\documentclass[12pt]{article}
\usepackage[dvips]{graphicx,color}
\usepackage{amsmath}
\usepackage{amsfonts}
\begin{document}
%%%%%%%%%%%%%%%%%%%%%%%%%%%%%%%%%%%%%%%%%%%%%%%%%%%%%%%%%%%%%%%%%%%%%
\title{Zero modes in non local domain walls}
\author{C.~D.~Fosco$^{a}$ and G.~Torroba$^{b}$\\
{\normalsize\it $^a$Centro At\'omico de Bariloche and Instituto Balseiro}\\
{\normalsize\it Comisi\'on Nacional de Energ\'\i a At\'omica}\\
{\normalsize\it 8400 Bariloche, Argentina}\\
{\normalsize\it $^b$Department of Physics and Astronomy}\\
{\normalsize\it Rutgers University}\\
{\normalsize\it Piscataway, NJ 08854, U.S.A.}\\}
\date{\today}
\maketitle
%====================================================================
\begin{abstract}
\noindent We generalize the Callan-Harvey mechanism to the case of
actions with a non local mass term for the fermions.  Using a
$2+1$-dimensional model as a concrete example, we show that both
the existence and properties of localized zero modes can also be
consistently studied when the mass is non local.  We derive some
general properties from a study of the resulting integral
equations, and consider their realization in a concrete example.
\end{abstract}
\bigskip
\newpage

%%%%%%%%%%%%%%%%%%%%%%%%%%%%%%%%%%%%%%%%%%%%%%%%%%%%%%%%%%%%%%%%%%%%%
%%%%%%%%%%%%%%%%%%%%%%%%%%%%%%%%%%%%%%%%%%%%%%%%%%%%%%%%%%%%%%%%%%%%%
The Callan-Harvey mechanism~\cite{Callan} explains the existence and
properties of fermionic zero modes that appear whenever the mass
of a Dirac field in $2k+1$ dimensions ($k=1,2,\cdots$) has a domain-wall
like defect. The zero modes due to this phenomenon are localized,
concentrated around the domain wall, and chiral from the point of view
of the domain-wall world-volume (a $2k$-dimensional theory).

This mechanism has found many interesting applications, both to
phenomenological issues~\cite{condensed} and theoretical
elaborations~\cite{fl,fls}.  Dynamical domain walls have been
considered in~\cite{fradkin}, and the case of a (dynamical)
supersymmetric model with this kind of configuration has been
discussed in~\cite{SUSY}.  Remarkably, lattice versions of the zero
modes, the so-called `domain wall fermions'~\cite{kap} have also been
precursors of the overlap Dirac operator~\cite{nar}, a sensible way to
put chiral fermions on a spacetime lattice.

In spite of the fact that the Callan-Harvey mechanism has been
extended in many directions, the existence of chiral zero modes
has been so far only studied for the case of {\em local\/} mass
terms, namely, those where the mass is a local function of the
spacetime coordinates. In this article, we relax that condition
and consider the case of a non local mass in $2+1$ dimensions.
This is a non trivial modification which does, however, arise
naturally in some applications. For example, when the fermionic
action is dressed by quantum effects, the corrections usually
adopt the form of non local form-factors.

Let us first review the standard Callan-Harvey mechanism~\cite{Callan}
when one considers a fermion field in 2+1 dimensions coupled to a
domain wall like defect, with the (Euclidean) action
\begin{equation}
  \label{eq:action1}
S({\bar\psi},\psi) \;=\; \int d^3x \,{\bar\psi}\,[\not\!\partial
\,+\, m (x) ] \,\psi \;.
\end{equation}
We use Euclidean coordinates  $x=(x_0 , x_1 , x_2)$ ($x_0$ is the Euclidean
time) and
$\not\!\partial= \gamma_\mu \partial_\mu\, ,$ where the $\gamma$-matrices are chosen
according to the convention:

\begin{equation}\label{dfgamma}
\gamma_0\;=\;\sigma_3 \;\;\;\gamma_1\;=\;\sigma_1
\;\;\;\gamma_2\;=\;\sigma_2\;.
\end{equation}
The \textit{local} mass $m(x)$ contains a topological defect; in the
simplest case of a rectilinear static defect~\cite{fl}, they have the
characteristic shape:
\begin{equation}\label{eq:aux}
m(x) \sim \Lambda \; \sigma(x_2) \; ,
\end{equation}
where $\sigma(x_2) \equiv \rm{sign} (x_2)$.  Therefore the domain wall, which is
the interface between two regions with different signs for $m(x)$, is
the $x_1$ axis.

From the Dirac operator
\begin{equation}
  \label{eq:defDirac1}
{\mathcal D}\;=\; \not \! \partial _x + m(x_2) \;\;,
\end{equation}
we can construct the hermitian operator $\mathcal H  = {\mathcal D}^{\dag} \,
\mathcal D$. The form of $\mathcal H$ suggests the introduction of the adjoint
operators
\begin{equation}
  \label{eq:defa1}
a\;=\; \partial_2+m(x_2) \;\;,\;\;\;
a^ \dag\;=\;-\partial_2+m(x_2)
\end{equation}
in terms of which
\begin{equation}
\mathcal H \,=\, (a^ \dag a -{\not \! {\hat \partial}}^2)P_L+(a a^ \dag-{\not \! {\hat \partial}}^2)P_R
\end{equation}
and
\begin{equation}
  \label{eq:Dirac1.1}
\mathcal D \,=\, (a+\not \! {\hat \partial})P_L+(a^ \dag+\not \! {\hat \partial})P_R
\end{equation}
where $P_L= \frac{1}{2} (1+\gamma_2)$, $P_R= \frac{1}{2} (1-\gamma_2)$.
Expanding $\psi(x)$ in the complete set of eigenfunctions of $a^ \dag a$ and
$a a^ \dag$, there appears~\cite{fl} a massless left fermion, localized
over the domain wall. Its $x_2$ dependence is dictated by the fact
that it is a zero mode of the $a$ operator, and it dominates the low
energy dynamics of the system.

We want to generalize this phenomenon to include a non local mass.
Namely, rather than (\ref{eq:defDirac1}), the Dirac operator shall
be
\begin{equation}
  \label{eq:defDirac2}
\tilde{\mathcal D}(x,y)\;=\; \not \! \partial _x  \; \delta (x-y)+ M(x,y) \; \; .
\end{equation}
Little can be said about the existence of a fermionic zero mode before
we make some hypotheses to restrict the form of $M(x,y)$. We assume
that the system has translation invariance in the coordinates $\hat{x}
\equiv (x_0,x_1)$ and that $M(x,y)$ consists of a local domain wall like
part, plus a non local term with a strength controlled by a
parameter $\lambda$:
\begin{equation}
  \label{eq:defnonlocalmass}
M(x,y)\;=\; m(x_2) \; \delta (x-y)- \lambda \; \int\frac{d^2{\hat k}}{(2\pi)^2} \,
e^{i {\hat k} \cdot ({\hat x} - {\hat y})} \gamma_k (x_2,y_2)  \;\;.
\end{equation}

We are looking for a zero mode $\Psi(x)$, so that:
\begin{equation}
  \label{eq:defzeromode1}
\langle x |\tilde{\mathcal D}|\Psi \rangle\;=\; \int \; d^3y \; \tilde{\mathcal{D}} (x,y)
\Psi(y) =0 \; \; .
\end{equation}
Taking advantage of the translation invariance in $\hat x$, we use
`separation of variables' to look for solutions of the form:
\begin{equation}
  \label{eq:defPsi1}
\Psi(x)\;=\; \chi(\hat x) \, \psi(x_2) \; \; ,
\end{equation}
where $\chi(\hat x)$ is a massless spinor, which is
left-handed from the point of view of the two-dimensional world
defined by ${\hat x}$, i.e.,
\begin{equation}\label{eq:propchi}
\not \! {\hat \partial}  \chi(\hat x) \;=\; 0 \;\;\;
P_R  \chi(\hat x) \;=\; 0\;.
\end{equation}
There is an essential difference regarding the space of solutions to
the equations above in Euclidean and Minkowski spacetimes. Indeed, in
Euclidean spacetime, it leads to analytic functions of $x_0+i x_1$,
while in the Minkowski case one has `left-mover' solutions.  Keeping
this distinction in mind, we continue working with the Euclidean
version.

Substituting (\ref{eq:defPsi1}) into (\ref{eq:defzeromode1}) and
comparing with (\ref{eq:Dirac1.1}), we arrive to a non local version
of the kernel for the annihilation operator
\begin{equation}
  \label{eq:nonlocala1}
a(x_2,y_2)\;=\;[\partial_2+m(x_2)] \, \delta(x_2-y_2)- \lambda \; \gamma_k
(x_2,y_2) \;\;.
\end{equation}
The zero mode $\psi(x_2)$, must then satisfy the equation
\begin{equation}
  \label{eq:zeromode2}
{\langle x |a|\psi \rangle\;}=\;[\partial_2+m(x_2)] \, \psi(x_2)- \lambda \;
\int_{-\infty} ^{+\infty}
dy_2 \, \gamma_k(x_2,y_2) \psi(y_2)=0 \; \; .
\end{equation}
Following the method of variation of parameters, we use the ansatz
\begin{equation}
  \label{eq:zeromode3}
\psi(x_2)\,=\, \psi_0(x_2) \varphi(x_2)
\end{equation}
where $\psi_0$ is the zero mode for the local part in
(\ref{eq:zeromode2}), which satisfies
\begin{equation}
[\partial_2+m(x_2)] \, \psi_0(x_2)\;=\;0 \;\;\;\;\; \psi_0(x_2)\,=\, N \,
exp[-\int^{x_2}_0 ds
\, m(s)] \;,
\end{equation}
and $N$ is a normalization constant.

The function $\varphi(x_2)$, which modulates $\psi_0(x_2)$, must
then satisfy the equation
\begin{equation}
\partial_2 \varphi(x_2)-\lambda \, \int_{-\infty}^{+\infty}
dy_2 \, [\psi_0 (x_2)^{-1} \, \gamma_k(x_2,y_2) \, \psi_0 (y_2)] \; \varphi (y_2)=0
\;.
\end{equation}
By integrating over $x_2$, the previous equation can be written in a
more convenient form as an integral equation
\begin{equation}
  \label{eq:fredholm1}
\varphi(x_2)-\lambda\,\int_{-\infty} ^{+\infty} dy_2 \,
\tilde{\gamma}_k(x_2,y_2) \varphi(y_2) = \varphi(0)
\end{equation}
where the kernel $\tilde{\gamma}_k$ is
\begin{equation}
  \label{eq:kernel1}
\tilde{\gamma}_k(x_2,y_2)\;=\; \int_0 ^{x_2} dz_2 \, \psi_0 (z_2)^{-1} \,
\gamma_k(z_2,y_2) \, \psi_0 (y_2)\;.
\end{equation}
This is a homogeneous integral equation, which can be conveniently
rewritten as an equivalent non-homogeneous set of equations. Indeed,
introducing linear operators, (\ref{eq:fredholm1}) can be rewritten as
follows:
\begin{equation}
  \label{eq:fredholm2}
(I-\lambda \, T) \varphi \;=\;c
\end{equation}
where
\begin{equation}
(I \varphi)(x_2) \equiv \varphi(x_2)\; , \;\;\; c \equiv \varphi(0) \;,
\end{equation}
\begin{equation}
  \label{eq:fredholm3}
(T \varphi)(x_2) \equiv \int dy_2 \; {\tilde \gamma}_k (x_2,y_2) \, \varphi(y_2) \;.
\end{equation}
We see that the problem can be discussed in terms of
(\ref{eq:fredholm2}), which is an inhomogeneous system of the Fredholm
type~\cite{fredholm}. The condition $c \equiv \varphi(0)$ has to be verified, of
course, after solving (\ref{eq:fredholm2}) for arbitrary $c$.

The reason for this procedure is that, had we used the original
homogeneous system, we should have had to introduce non compact
operators, and the theory for this kind of operator is much
poorer than for the compact case.

Physical restrictions imposed on the non local mass naturally lead us
to integral equations of the Fredholm type~\cite{fredholm}. The
`Fredholm alternative'~\cite{fredholm} states that, if $A=I-\lambda \, T$,
where $T$ is a compact operator on a Hilbert space $H$, then the
following alternative holds: (a) either $A \varphi_0 =0$ has only the
trivial solution, in which case $A \varphi = c$ has a unique solution $\forall c \in
H$; or (b) $A \varphi_0 =0$ has $q$ linearly independent solutions $\varphi_i \, \in
H$.  Then $A^\dagger {\tilde \varphi}_0 =0 $ also has $q$ linearly independent
solutions ${\tilde \varphi}_i \, \in H$. In this case $A \varphi=c$ is solvable iff
$(c,{\tilde \varphi}_i)=0 \;\; \forall i=1, \dots, q$.

In the case (b), the general non-homogeneous solution is
\begin{equation}
\varphi=\varphi_p+\sum_{i=1}^q a_i \varphi_i
\end{equation}
where $\varphi_p$ is a particular solution and $a_i$ are arbitrary
constants.

On the other hand, when the alternative (a) holds true, this
implies that the solution shall be unique when $c$ is replaced by
$\varphi(0)$. Solutions corresponding to $c \neq \varphi(0)$ are
not solutions of the system: \mbox{$(I-\lambda \, T) \varphi =
c$}, \mbox{$c \equiv \varphi(0)$}, equivalent to the original
homogeneous equation (\ref{eq:fredholm1}), and may, therefore, be
safely discarded.

Any true solution of the system will also verify a subsidiary
equation, obtained by setting $x_2 = 0$ in (\ref{eq:fredholm1}):
\begin{equation}
\int_{-\infty} ^{+\infty} dy_2 \,\tilde{\gamma}_k(0,y_2) \varphi(y_2) = 0
\end{equation}
for  any $\lambda \neq 0$. This equation  shall be  true whenever the equations
$(I-\lambda \, T) \varphi = c$, $c \equiv \varphi(0)$ are both true (since it is derived from
them), and any solution will automatically verify it.

In our case, we want to study the effect of the non local term, the
strength of which is controlled by the value of $\lambda$. Close enough to the
local mass case, $\lambda$ can be made arbitrarily small, so
$\lambda^{-1}$ is not an eigenvalue of $T$ and consequently $A \varphi=0$ has
only the trivial solution. Therefore, if $T$ is a Fredholm operator,
(\ref{eq:fredholm1}) will have a unique solution. The functional space $H$ to
which $\varphi(x)$ belongs is restricted by the condition
\begin{equation}\label{eq:hcond}
\int dx_2 \; \Big(\psi_0(x_2) \Big)^2 \Big(\varphi(x_2)\Big)^2 < \infty  \; ,
\end{equation}
because $\psi(x)$ itself has to be normalizable.  This becomes a Hilbert
space, and it contains the zero mode of the local operator ($\varphi\,=\,
{\rm constant}$), when the scalar product used in $H$ is the one
defined by (\ref{eq:hcond}), namely,
\begin{equation}\label{eq:defscalar}
( f, g ) \;\equiv\; \int dx_2 \,\Big(\psi_0(x_2) \Big)^2 [f(x_2)]^* \, g(x_2) \;.
\end{equation}

When Fredholm's hypotheses are satisfied for a general kernel
$\tilde{\gamma}_k(x_2,y_2)$, it is possible to find a perturbative
solution. Indeed, writing
\begin{equation}
\varphi(x_2)\;=\;\varphi(0)+\lambda\,\int_{-\infty} ^\infty dy_2 \,
\tilde{\gamma}_k(x_2,y_2) \, \varphi(y_2) \;,
\end{equation}
and successively replacing $\varphi(y_2)$ by $\varphi(0)+\lambda\,\int_{-\infty} ^\infty
dy_2 \,
\tilde{\gamma}_k(x_2,y_2) \varphi(y_2)$ on the right hand side, we obtain the Neumann
series~\cite{fredholm}
\begin{equation}
\varphi(0)+\lambda \, K_1 (x_2) \, \varphi(0)+\dots+\lambda^n \, K_n (x_2)\, \varphi(0)+\dots
\end{equation}
where
\begin{equation}
  \label{eq:kernelj}
K_j (x_2) = \int _{-\infty} ^\infty \, dy_2 \, \gamma^{(j)}(x_2,y_2)
\end{equation}
and
\begin{eqnarray}
\gamma^{(1)}(x_2,y_2)&=& \tilde{\gamma}_k (x_2,y_2)\;\;,\;\;
\gamma^{(2)}(x_2,y_2)\;=\; \int_{-\infty}^\infty \, dz_2 \; \tilde{\gamma}_k (x_2,z_2)
\tilde{\gamma}_k (z_2,y_2)\nonumber\\
& \ldots & \gamma^{(n+1)}(x_2,y_2)\;=\; \int_{-\infty}^\infty \, dz_2 \; \tilde{\gamma}_k (x_2,z_2)
\gamma^{(n)} (z_2,y_2) \;.
\end{eqnarray}
It can be shown~\cite{fredholm} that the Neumann series converges
uniformly to the solution $\varphi(x)$ when
\begin{equation}
  \label{eq:radiusconv}
|\lambda|<\frac{1}{\mu} \; ,
\end{equation}
\begin{equation}
  \label{eq:mucuad}
\mu^2=\int_{-\infty}^\infty \, dx_2 \, dy_2 \; [\tilde{\gamma}_k (x_2,y_2)]^2 \; .
\end{equation}
So choosing $\lambda$ to satisfy (\ref{eq:radiusconv}), we are allowed to represent
$\varphi$ by the expansion
\begin{equation}
  \label{eq:neumann}
\varphi(x_2)=\varphi(0)+\lambda \, K_1 (x_2) \, \varphi(0)+\dots+\lambda^n \,
K_n (x_2)\, \varphi(0)+\dots
\end{equation}

%%%%%%%%%%%%%%%%%%%%%%%%%%%%%%%%%%%%%%%%%%%%%%%%%%%%%%%%%%%%%%%%%%%%%
%%%%%%%%%%%%%%%%%%%%%%%%%%%%%%%%%%%%%%%%%%%%%%%%%%%%%%%%%%%%%%%%%%%%%
To make the previous analysis more concrete, we consider now an
example, based on a specific choice of both the local and non local
parts of $M(x,y)$ in (\ref{eq:defnonlocalmass}).  A natural
generalization of the purely local mass case is to have \mbox{$m(x_2)=
  \Lambda \, \sigma(x_2)$} and a $\gamma_k(x_2,y_2)$ which is `strongly diagonal' and
symmetric in $(x_2,y_2)$, i.e.:
\begin{equation}
  \label{eq:kernelspecific}
\, \gamma_k(x_2,y_2) \,=\,\frac{1}{2} \,
\Big[\sigma(x_2)+\sigma(y_2)\Big]\delta_N (x_2-y_2) \;,
\end{equation}
where $\delta_N$ is an approximation of Dirac's delta: $\delta_N (x_2-y_2) \to
\delta(x_2-y_2)$ when $N\to\infty$. Note that $\gamma_k(x_2,y_2) \to {\rm sign}(x_2) \, \delta(x_2-y_2)$ when
$N\to\infty$, so that the non local term reduces to a local domain wall mass.

We adopt `natural' units such that $\Lambda= 1$, and study the particular
case $|x_2|\leq L = \Lambda^{-1}$. Regarding $\delta_N$, we use a truncation of the
one-dimensional completeness relation:
\begin{equation}
\delta_N (x_2,y_2)\,=\,\sum_{n=0}^N \, \varphi_n(x_2) \, \varphi_n
^{\dag}(y_2) \;,
\end{equation}
where $\{\varphi_n \}$ is a complete set of functions. We chose $\varphi_n(x)$ to
be the harmonic oscillator's eigenfunctions.

The normalized zero mode corresponding to the local part is
\begin{equation}
  \label{eq:zeromodelocal2}
\psi_0 (x_2) \,=\,N_0 \, \exp \Big[-\int_0 ^{x_2} dt \, m(t)
\Big]\,=\,\frac{1}{\sqrt{1-e^{-2}}} \, e^{-|x_2|} \; ,
\end{equation}
and the corrected zero mode $\psi(x_2)\,=\,\psi_0 (x_2) \, \varphi(x_2) \;$ is
then determined by $\varphi(x_2)$. The integral equation for $\varphi(x_2)$
becomes
\begin{equation}
  \label{eq:integrequat}
\varphi(x_2)\,=\,\varphi(0)+\lambda \, \int_{-1}^1 dy_2 \;
\tilde{\gamma}(x_2,y_2) \, \varphi(y_2)
\end{equation}
where
$$
\tilde{\gamma}(x_2,y_2) \,=\,\frac{1}{2} \int_0 ^{x_2} \,dz_2 \; e^{|z_2|}
\, \Big[\sigma (z_2) \, + \, \sigma (y_2) \Big] \,
$$
\begin{equation}
  \label{eq:gammacompact}
\times \Big[ \sum_{n=0}^N \, \varphi_n(z_2) \, \varphi_n^{\dag}(y_2) \Big ] \, e^{-|y_2|} \; .
\end{equation}

\begin{figure}
\begin{center}
\begin{picture}(0,0)%
\includegraphics{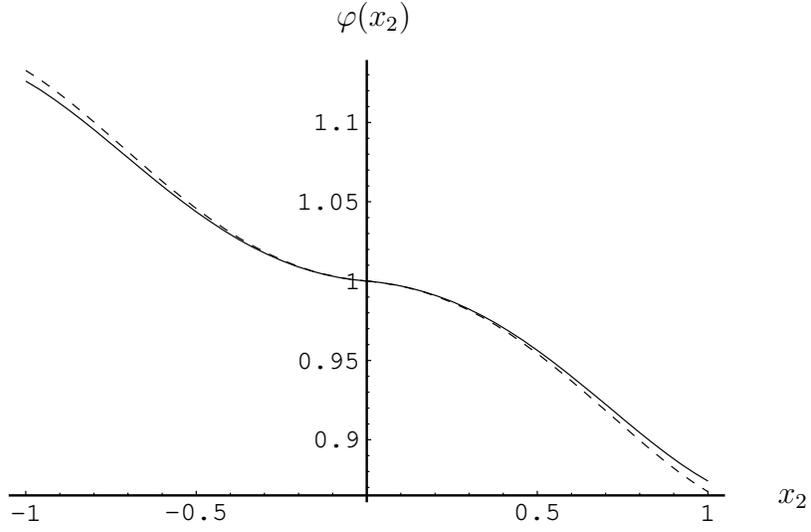}%
\end{picture}%
\setlength{\unitlength}{3947sp}%
\begingroup\makeatletter\ifx\SetFigFont\undefined%
\gdef\SetFigFont#1#2#3#4#5{%
  \reset@font\fontsize{#1}{#2pt}%
  \fontfamily{#3}\fontseries{#4}\fontshape{#5}%
  \selectfont}%
\fi\endgroup%
\begin{picture}(4855,3254)(1596,-12246)
\put(3676,-9136){\makebox(0,0)[lb]{\smash{\SetFigFont{12}{14.4}{\rmdefault}{\mddefault}{\updefault}{\color[rgb]{0,0,0}$\varphi(x_2)$}%
}}}
\put(6451,-12136){\makebox(0,0)[lb]{\smash{\SetFigFont{12}{14.4}{\rmdefault}{\mddefault}{\updefault}{\color[rgb]{0,0,0}$x_2$}%
}}}
\end{picture}
\end{center}
\caption{\footnotesize{$\varphi(x_2)\,$ after two iterations of the Neumann series
(\ref{eq:neumann}) for the $x_2$-compactified case. The dashed line corresponds
to the first iteration.}}
\label{varphicompact}
\end{figure}

From equations (\ref{eq:radiusconv}) and (\ref{eq:mucuad}), we see
that the Neumann series converges, in this case, for $|\lambda|<0.0990$. So,
we assume that $\lambda=0.01$, and calculate (\ref{eq:integrequat})
perturbatively. Since its expression in terms of analytic functions is
not very illuminating, we present, in Figure~\ref{varphicompact}, the
numerical results of the first two iterations, taking $\varphi(0)=1\,$ and
$N=3\,$.

\begin{figure}
\begin{center}
\begin{picture}(0,0)%
\includegraphics{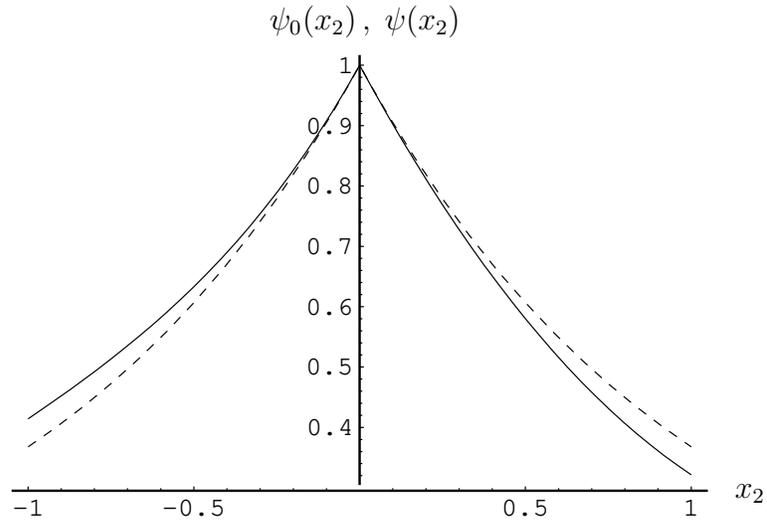}%
\end{picture}%
\setlength{\unitlength}{3947sp}%
\begingroup\makeatletter\ifx\SetFigFont\undefined%
\gdef\SetFigFont#1#2#3#4#5{%
  \reset@font\fontsize{#1}{#2pt}%
  \fontfamily{#3}\fontseries{#4}\fontshape{#5}%
  \selectfont}%
\fi\endgroup%
\begin{picture}(4567,3199)(1659,-12246)
\put(6226,-12136){\makebox(0,0)[lb]{\smash{\SetFigFont{12}{14.4}{\rmdefault}{\mddefault}{\updefault}{\color[rgb]{0,0,0}$x_2$}%
}}}
\put(3301,-9191){\makebox(0,0)[lb]{\smash{\SetFigFont{12}{14.4}{\rmdefault}{\mddefault}{\updefault}{\color[rgb]{0,0,0}$\psi_0 (x_2) \, , \; \psi(x_2)$}%
}}}
\end{picture}
\end{center}
\caption{\footnotesize{Zero mode profiles for  local (dashed line) and
    non local (full line) masses.}}
\label{compactmodes}
\end{figure}

In Figure~\ref{compactmodes}, we show the original zero mode
(\ref{eq:zeromodelocal2}) and the one including order-$\lambda^2$
corrections. We see that the corrected zero mode continues to be
localized over the domain wall, although it is no longer a symmetric
function of $x_2$.  Besides, this perturbative method introduces only
smooth corrections in the zero mode, as expected.

%%%%%%%%%%%%%%%%%%%%%%%%%%%%%%%%%%%%%%%%%%%%%%%%%%%%%%%%%%%%%%%%%%%%%
%%%%%%%%%%%%%%%%%%%%%%%%%%%%%%%%%%%%%%%%%%%%%%%%%%%%%%%%%%%%%%%%%%%%%
%%%%%%%%%%%%%%%%%%%%%%%%%%%%%%%%%%%%%%%%%%%%%%%%%%%%%%%%%%%%%%%%%%%%%

We conclude by summarizing that we have studied a generalization of
the Callan-Harvey mechanism to the case of a non local mass in 2+1
dimensions, showing that for a certain set of assumptions about the
non locality, there continues to exist a fermionic zero mode.

We have first considered a quite general non local term, deriving a
linear integral equation for a function which modulates the zero mode
of the local case, and accounts for the effect of the non local domain
wall.  Considering a defect of finite size, the Fredholm alternative
theorem applies and there is a unique, localized chiral zero
mode. Moreover, perturbation theory may be applied to calculate it:
for the concrete example of a strongly diagonal mass, we have calculated
the first few terms in a perturbative expansion, showing that they lead to
negligible modifications with respect to the local mass case.

\section*{Aknowledgments}
G.~T.\ is supported by Rutgers Department of Physics. C.~D.~F.\ is
supported by CONICET (Argentina), and by a Fundaci\'on Antorchas
grant.

\end{document}